# Application of a spring-dashpot system to clinical lung tumor motion data

**E J Ackerley[1], A E Cavan[2,3], P L Wilson[1†], R I Berbeco[4] and J Meyer[2,5,‡]**

[1]Department of Mathematics and Statistics, University of Canterbury, Private Bag 4800, Christchurch 8140, New Zealand

[2]Department of Physics and Astronomy, University of Canterbury, Private Bag 4800, Christchurch 8140, New Zealand

[3]Department of Medical Physics & Bioengineering, Christchurch Hospital, Private Bag 4710, Christchurch, New Zealand

[4]Department of Radiation Oncology, Brigham and Women's Hospital, Dana-Farber Cancer Institute and Harvard Medical School, Boston, Massachusetts 02115, USA

[5]Department of Radiation Oncology, University of Washington Medical Center, Box 356043, Seattle, Washington 98195-6043, USA

[†] Corresponding author: Department of Mathematics and Statistics, University of Canterbury, Private Bag 4800, Christchurch 8140, New Zealand, Phone: +64 3 364 2664, Fax: +64 3 364 2587, Email: phillip.wilson@canterbury.ac.nz

[‡] Present address: Department of Radiation Oncology, University of Washington, University of Washington Medical Center, 1959 NE Pacific St, Box 356043, Seattle WA 98195

## Abstract

A spring-dashpot system based on the Voigt model was developed to model the correlation between abdominal respiratory motion and tumor motion during lung radiotherapy. The model was applied to clinical data comprising 52 treatment beams from 10 patients, treated on the Mitsubishi Real-Time Radiation Therapy system, Sapporo, Japan. In Stage 1, model parameters were optimized for individual patients and beams to determine reference values and to investigate how well the model can describe the data. In Stage 2, for each patient the optimal parameters determined for a single beam were applied to data from other beams to investigate whether a beam-specific set of model parameters is sufficient to model tumor motion over a course of treatment.

In Stage 1 the baseline root mean square (RMS) residual error for all individually-optimized beam data was 0.90 ± 0.40 mm. In Stage 2, patient-specific model parameters based on a single beam were found to model the tumor position closely, even for irregular beam data, with a mean increase with respect to Stage 1 values in RMS error of 0.37 mm. On average the obtained model output for the tumor position was 95% of the time within an absolute bound of 2.0 mm and 2.6 mm in Stage 1 and 2, respectively.

The model was capable of dealing with baseline, amplitude and frequency variations of the input data, as well as phase shifts between the input tumor and output abdominal signals. These results indicate that it may be feasible to collect patient-specific model parameters during or prior to the first treatment, and then retain these for the rest of the treatment period. The model has potential for clinical application during radiotherapy treatment of lung tumors.

## 1. Introduction

Lung cancer was the most commonly diagnosed cancer worldwide in 2008, with an incidence rate of over 1.5 million new cases, or 12 % of the total new cancer diagnoses[1]. It was also the most common cause of death from cancer, accounting for 17% of all cancer deaths. Radiation therapy is a common modality for lung cancer treatment, however treatment efficacy is limited by the motion of the lungs during respiration[2], which is primarily driven by diaphragm motion, and to a lesser extent by chest motion. The magnitude of lung tumor motion depends on patient-specific breathing and tumor characteristics, and is usually most pronounced along the superior-inferior (SI) axis, compared to the anterior-posterior (AP) and lateral directions[3]. One study, which included 39 patients treated with the Real-Time Radiation Therapy system in Sapporo, Japan, found a median tumor movement of 1.1 mm, 2.3 mm and 5.4 mm in lateral, AP and SI direction, respectively[4].

Typically, for lung tumors the clinical target volume (CTV) is enlarged to the internal target volume (ITV) with the intent of ensuring sufficient tumor coverage in the presence of motion[5]. This strategy can result in excessive irradiation of surrounding healthy tissue, or marginal miss of the tumor[6]. The high doses required for tumor control are close to or above the tolerance level for healthy tissues, resulting in increased side effects, or requiring a reduction in dose to the tumor, decreasing tumor control probability (TCP)[7, 8]. In this way, compensating for the presence of respiratory motion of lung tumors can improve the therapeutic ratio and thus survival rates[9].

One means to compensate for tumor motion is by means of real-time tumor tracking and motion compensation[8, 10-14]. Motion tracking is complicated by variations in baseline, frequency and oscillatory amplitude and form, despite the overall superficial regularity of respiratory motion. These effects can differ widely between patients but can also vary over the course of treatment of a single patient[7]. Methods to suppress such variability include abdominal compression to reduce tumor mobility, breath control techniques (active or passive) and respiratory gating of irradiation[3, 7, 15-19], with the latter usually based on external surrogate motion. All of these techniques have some limitations, such as the need for active patient cooperation, consistent ability to maintain total lung capacity[20], or a lengthened treatment time for gated therapy due to the required beam-off periods.

Modeling of the respiratory-induced lung tumor motion can facilitate dynamic tracking and compensation for real-time and gated treatments[4, 7, 8, 19]. Direct tumor tracking systems may use portal imaging[10, 13, 21, 22] or implanted fiducial markers in the tumor, in conjunction with a diagnostic x-ray imaging system[20, 23, 24], but this continuous imaging can impart a considerable radiation dose which is not always clinically justifiable[25].

By contrast, indirect tumor tracking systems use internal or external surrogates to obtain tracking signals. A model is then required to relate the surrogate to tumor motion. Examples of external respiratory surrogates are a spirometer, strain gauge or abdominal markers[20], while an internal surrogate can be the diaphragm motion[26]. Such indirect approaches require the relationship between surrogate and internal tumor motion to be consistent and well correlated[27]. The benefits of indirect tracking include the elimination of risks associated with changes in the relationship between tumor and implanted marker during radiotherapy[4], fiducial implantation, extra radiation doses, and direct tracking failures. Most models correlating surrogate data to tumor motion have been so-called ‘black box’ or ‘grey box’ approaches, for which the internal behavior of the physical system is not or only partially known, respectively. These models consider the input and output data, but not the exact physical relationship between them. Various approaches of this form include linear correspondence models[9, 28], composites of baseline drift, frequency variation, fundamental pattern change and random observation noise[7], characterization of the motion with a piecewise linear model of defined stages of the breathing cycle[29], state-based probabilistic models[30], least-squares parameter estimation and systems identification[3] or adaptive neural networks[31]. These models generally require a large amount of sample data which encompasses the range of possible relationships between the input and output states. Many of the previous external surrogate models tend to deal poorly with irregularities in the breathing pattern, such as baseline drift or a hysteresis[31], a lag or phase shift between internal and external motions[27, 32, 33], and usually have a strong dependence on tumor and marker locations, motion dimensions and type of breathing pattern[3, 4, 28]. One of the more promising models is the model by Cervino *et al.*[26] using an internal surrogate signal of the diaphragm; however the practical benefits of using an external signal warrant further investigation into external abdominal surrogates.

A more physical approach to the problem models lung motion as a contact problem of elasticity theory by describing the physiology of breathing using elastic constants to directly model the lung tissue[34]. A different approach was explored by Wilson and Meyer[35, 36] who presented a comprehensive physical 3D system of springs and dashpots to model the correlation of an external abdominal respiratory signal and the lung tumor motion, rather than directly modeling the actual lung tissue[24]. Wilson and Meyer showed mathematically that it is possible to formally simplify the three-dimensional model to a one-dimensional model when SI motion of the tumor dominates.

To overcome some of the physical intrinsic limitations for the practical application of the above approach, the first aim of this paper was to refine the Wilson and Meyer model in the one-dimensional realm. The relationship of lung tumor and external abdominal marker movement is considered for the primary dimension of motion, i.e. the SI and AP direction of the tumor and abdominal signal, respectively. The second objective was to apply the refined model to a large and realistic clinical data set of tumor motion from a cohort of lung cancer patients treated with the Mitsubishi Real-Time Radiation Therapy system in Sapporo, Japan to investigate the model behavior on an intra- and inter-patient specific basis.

## 2. Material and Methods

### *2.1 Mathematical model*

Following the methodology of Wilson & Meyer[35], we model the correlation between the surrogate data provided by abdominal motion and the target data of the lung tumor motion with a pseudo-mechanical system. This system is composed of springs and dashpots, the latter providing a damping effect. In general, a spring-dashpot unit can be composed of springs and dashpots in series, parallel, or some combination thereof. Three of the most frequently-used configurations in other applications are shown in Figure 1. The full three-dimensional system of Wilson & Meyer (2009) featured a Voigt unit able to act in each Cartesian direction of motion[1].

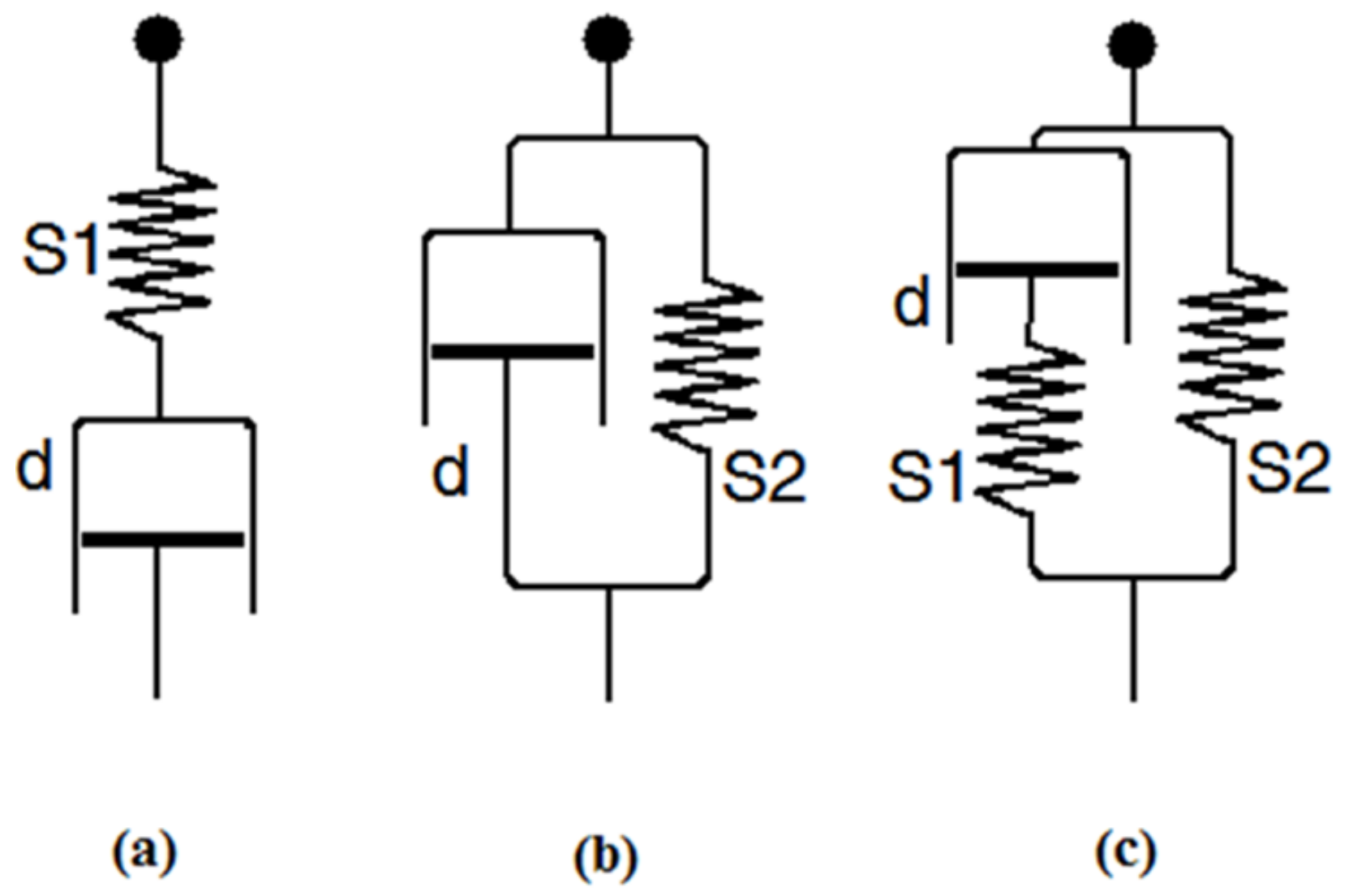

***Figure 1:*** (a) Maxwell model (b) Voigt model (c) Standard Linear Solid model. *S* denotes a spring and *d* denotes a dashpot.

Asymptotic analysis[35] shows that when the SI motion of the tumor is dominant the motion is well-modeled by a single Voigt unit acting in the SI direction, with the tumor attached at one end and the abdominal system providing input at the other. It is this pseudo-one dimensional class of tumor motion which we consider in this paper. In the following condensed derivation, all variables and parameters are non-dimensional.

[1] Note that these were incorrectly drawn in the Wilson & Meyer paper as Maxwell units.

The variable *x* is the tumor displacement from a resting equilibrium position in the SI direction. The Hookean spring has associated parameter $\omega$, while the dashpot, characterized by the parameter $\lambda$ provides a retarding force directly proportional to the tumor velocity. Although the dominant time-dependent abdominal motion, *f(t)*, is in the AP direction perpendicular to the dominant tumor motion direction, this component of the abdominal signal is linearly scaled and used as input to the model. This forcing input signal, $x^*(t)$, is therefore related to the measured data by $x^* = \delta f$, with the scale factor $\delta$ being determined under optimization. This approach reduces the dimension of the parameter space of Wilson & Meyer by one, and also allows a wider range of parameter values to be studied, both of which enable the model to be more readily optimized.

Following Wilson & Meyer[35] but with the above notation, the tumor motion is given by

$$\frac{d^2x}{dt^2} = -2\lambda\frac{dx}{dt} - \omega^2 x + \delta f \tag{1}$$

The model will be shown to hold for all patients in this class of tumor motion, with values of the parameter triplet $(\omega, \lambda, \delta)$ derived from a "training" subset of patient data, and successfully applied to model the remaining data.

*2.2 Numerical approach*

Equation (1) was rewritten as two coupled first-order ordinary differential equations, which were solved for each set of training data using the MATLAB™ differential equation solver *ode45*, which is recommended for non-stiff problems[37]. The resulting output data was optimized with respect to the model parameter triplet $(\omega, \lambda, \delta)$ using the MATLAB *fminsearch* routine in combination with a cost function: an iterative process used to find the minimum of the root mean squared error (RMS) (2) between the measured tumor position data, $X$, and the model output values, $x$ (at each time step, $n$).

$$RMS = \left( \frac{\sum_{i=1}^{n} (X - x)^2}{n} \right)^{\frac{1}{2}} \qquad (2)$$

The algorithm starts with an initial estimate of the triplet based on the model characterization in Wilson & Meyer and proceeds via a process of unconstrained non-linear optimization with a Simplex search method. Optimization is initially patient- and beam-specific and is based on a set of “training” data constituting approximately 2 minutes of patient data. The algorithm terminates when the RMS error is less than $10^{-3}$ or when a maximum number of 100 iterations have been reached.

In *Stage 1* of our work, optimized parameters obtained in this way are used to model the tumor motion via numerical solutions to equation (1) for each beam to obtain reference values. We refer to this stage as ‘optimized’.

In *Stage 2* ( ‘non-optimized’) we take the optimized parameters from a single beam – e.g. Day 1, Beam 1– and use these to model the tumor motion via (1) for that patient's respiratory data obtained from other beams on the same day and consecutive days. The comparison between the predicted tumor position, $x$, and the clinical data, $X$, is presented and discussed in sections 3 and 4.

### *2.3 Clinical data*

The clinical data consists of 3D internal fiducial-based motion obtained using radiopaque fiducial markers implanted and visualized in real-time using stereoscopic diagnostic x-ray fluoroscopy, collected using a Mitsubishi Real-Time Radiation Therapy system, at the Nippon Telegraph and Telephone Corporation Hospital, Sapporo, Japan[27]. This was obtained simultaneously with 1D external abdominal motion, collected on an independent co-ordinate system using a laser based AZ-733V "RespGate" made by Anzai Medical, Tokyo, Japan. The data set consisted of motion results from eleven patients (patient 1-11) with lung cancer at various sites, with a total of 171 treatment beams for up to ten consecutive treatment fractions (day 1-10) and four beam configurations per fraction (beam 1-4) [(2)]. The patients were not a random sample of the general lung cancer population, but were selected on the basis of an estimated internal marker motion of more than 10 mm peak to peak (SI, lateral or AP direction). The data show phase shifts of the internal-external data (one signal lagging the other)[32, 33] in the SI direction that are mostly between 100 and 200 ms.

The data were not acquired in a common coordinate system with absolute spatial co-ordinates and therefore they could not be normalized to a common reference point. Spatial baselines for the tumor and abdominal markers were defined by normalizing on the mean of each beam data.

For baseline values in Stage 1 the model was initially applied to data from 52 treatment beams from 10 patients: beam 1 data for all days for all patients (as this type of data was available for all patients), as well as all beams for patients 1-3. More detailed analysis was then carried out on representative individual patients: one whose model performance was close to the mean of all patients (Patient 7); one with relatively poor model performance (Patient 8); and one with excellent model performance (Patient 10).

In the second stage, for each patient the optimal parameters determined for an arbitrarily chosen single beam were applied to data from other beams and treatment days to investigate whether a beam-specific set of model parameters is adequate to model tumor motion over a course of treatment.

---

[(2)] Patient 5 had eight fractions, then a further four two months later. The analysis was kept consistent. The data for patient 11 were unreliable and not included.

We also investigate the ability of the model to adapt to irregularities in the clinical data such as unclear tracking data, missing data points, baseline drift and amplitude and frequency variation.

## 3. Results

### *3.1 Stage 1*

Applying the optimization algorithm to the data from the 52 clinical treatment beams with the initial estimate parameter triplet ($\omega$, $\lambda$, $\delta$) = (10, 5, 200) gave, on average, a root mean square error of 0.90 mm (2 s.f.) and a standard deviation of 0.40 mm between the modeled position and the tumor motion in the superior-inferior direction. The best result from all the beam data modeled, Patient 4, Day 1, Beam 1, had an RMS error of 0.38 mm. An overview of the RMS errors for all patients is shown in Figure 2.

The number of iterations required for optimization varied considerably with each beam but was generally less than 100 iterations. A representative example (patient 7) of beam data and the model results with a RMS of 0.69 mm is shown in Figure 3, with optimized parameters $\omega = 10.12$, $\lambda = 5.27$, $\delta = 197.91$.

The linear relationship between the position predicted by the model and the actual tumor location was investigated, and the regression correlation coefficient was calculated for each beam studied. The mean correlation from all 52 beams was 0.96 with standard deviation of 0.03, which would indicate the existence of strong linear relationships. A graphical analysis of one beam each for the three representative patients is shown in the top row of Figure 4.

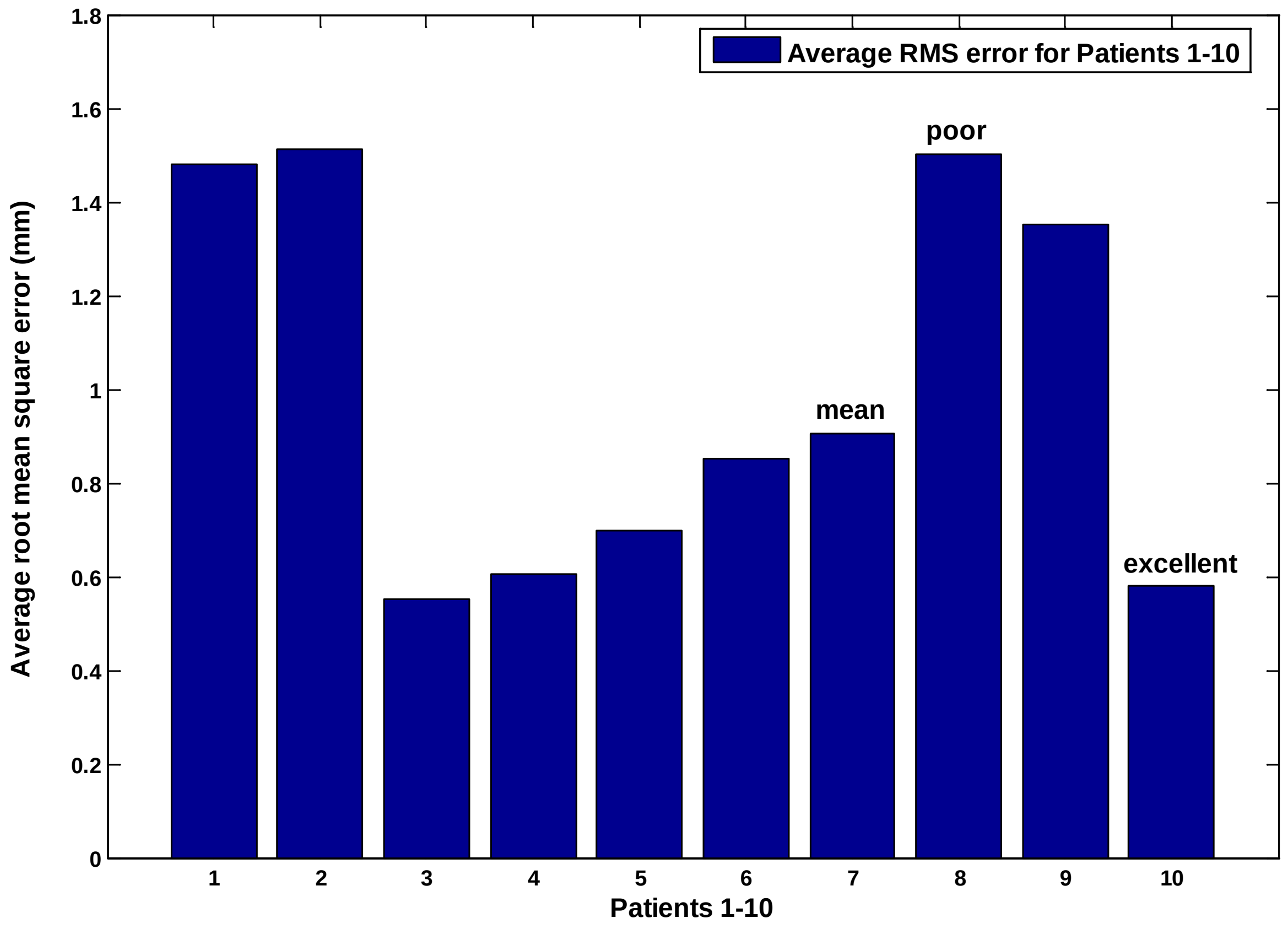

***Figure 2:*** Average root mean square error (mm) between model output and clinical data for each patient. Representative patients 7, 8 and 10 used for further analysis are marked as ‘mean’, poor’ and ‘excellent’, respectively. Note that the number of beams included to calculate the RMS errors varies between patients.

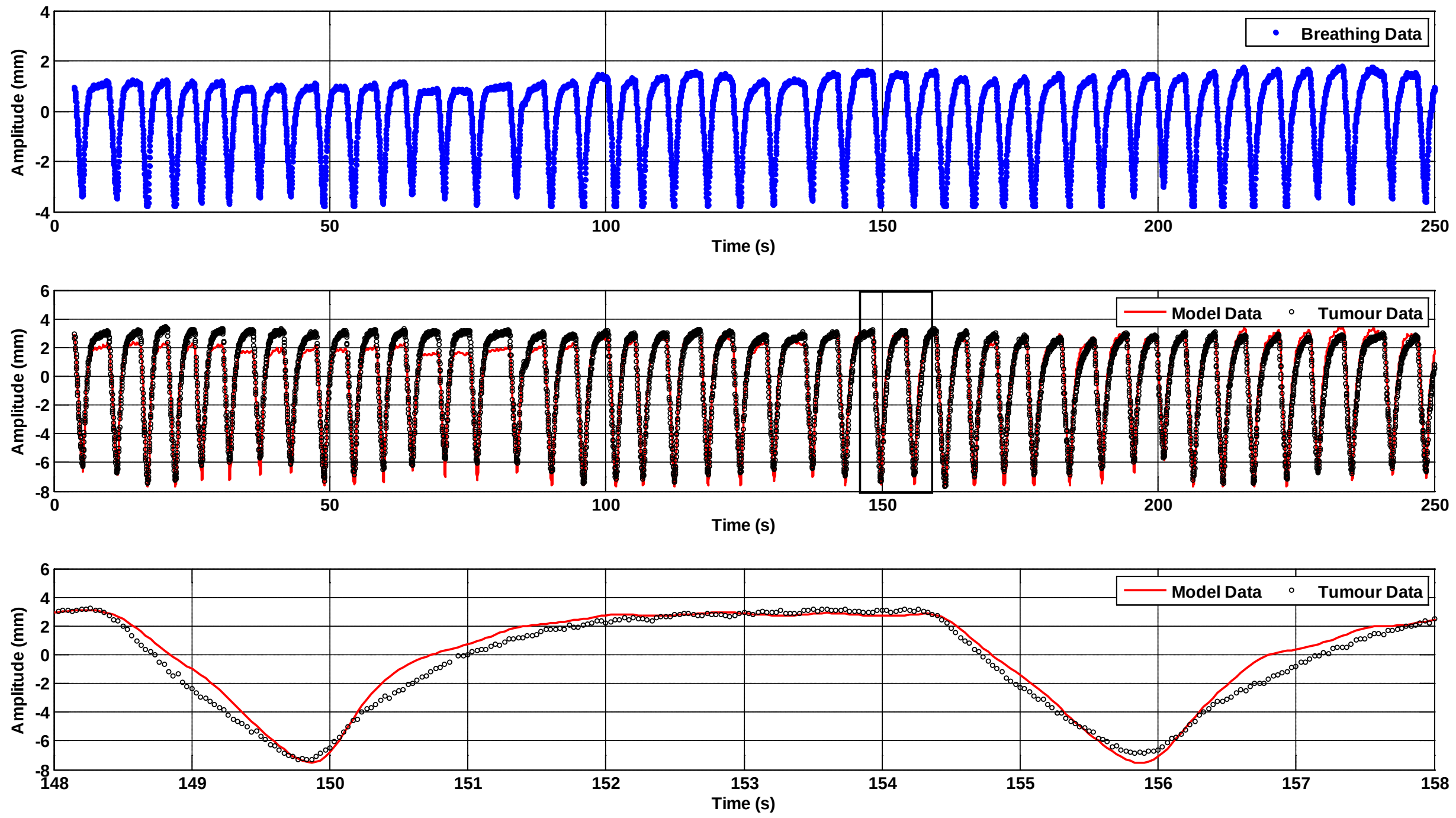

***Figure 3:*** The clinical data and model output, showing breathing pattern, tumor motion, and model output for Patient 7, Day 4, Beam 1. The lower figure shows a close-up view of part of the tumor data and model output.

Additionally, we considered the residual vectors for each beam, which appeared visually to have a normal-like distribution, although a more rigorous testing of these vectors revealed that only four beams were sufficiently normal to satisfy the Jarque-Bera[38] test. The bottom row of Figure 4 shows the histograms of three beams. Note that the best fitting probability distribution is scaled to the area of the histograms. The full width at half maximum was 1.6 mm, 3.9 mm and 1.2 mm for patients 7, 8 and 10, (mean, poor, excellent), respectively.

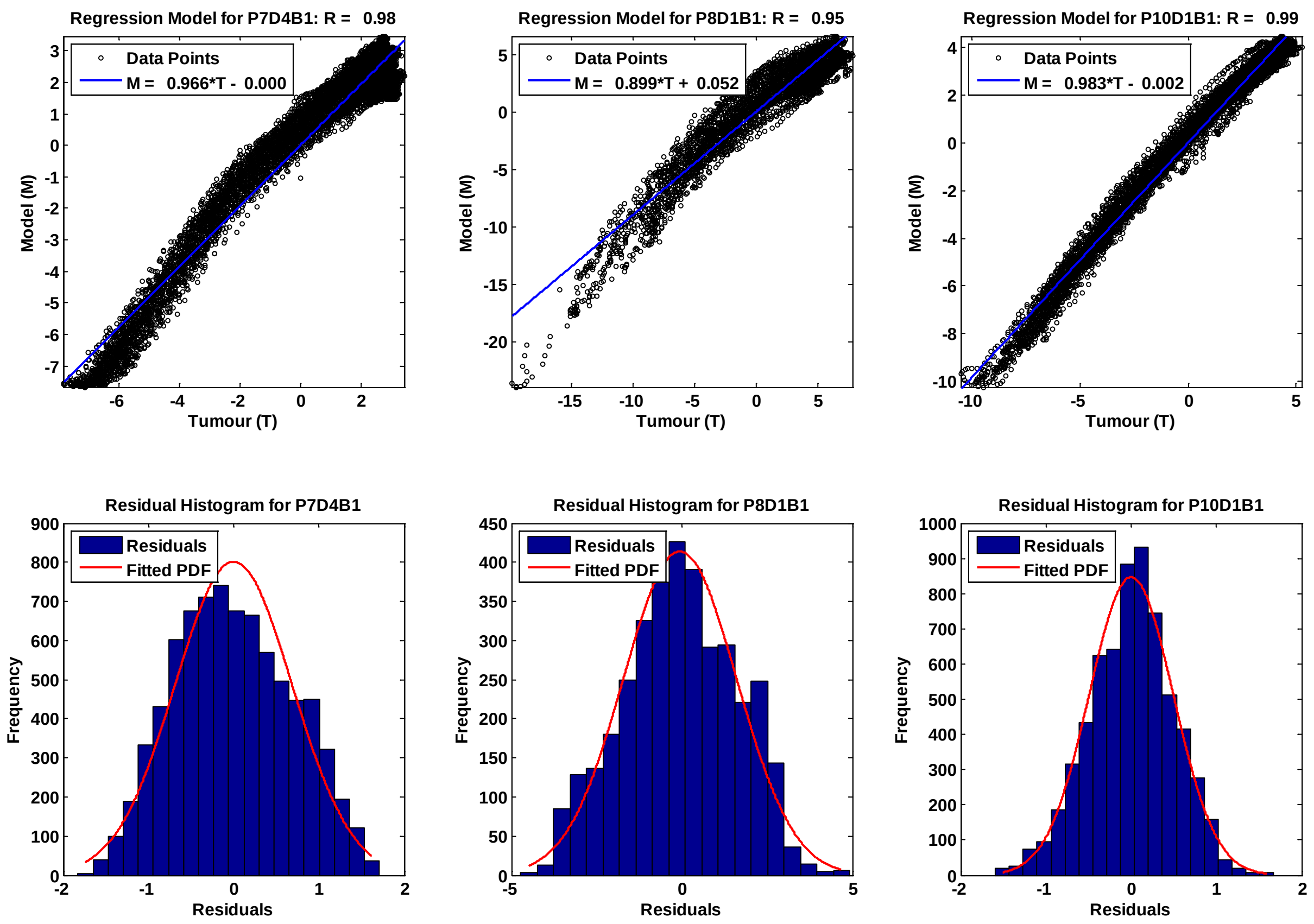

***Figure 4:*** Regression plots and residual histograms for three beams: “mean” Patient 7, Day 4, Beam 1 (left); “poor” Patient 8, Day 1, Beam 1 (center); “excellent” Patient 10, Day 1, Beam 1 (right).

For clinical applications, a major concern is the proportion of time, which we term the *residency,* for which the model output is within a small distance of the actual tumor position, i.e. within the treatment margin. The residency of the 52 beams was calculated for bounds on the absolute difference in positions of 1, 2, 3, 4, 5 mm. The residencies from Patients 7, 8 and 10, and all 52 beams are shown in Table 1. As can be seen from the table, to be resident 95% of the time a bound of 2 mm is required on average. In the second stage the parameter results obtained from optimizing the Day 1, Beam 1 data for each patient were applied to other beams and days for that patient in lieu of optimizing the data from each beam independently (Stage 1). There was an average increase in RMS error of 0.37 mm. Results are shown in Table 2.

When the robustness of the model was examined with respect to the linear relationship between the tumor position and model prediction, the mean correlation coefficient of the 52 beams was 0.96, the same value as when each beam was optimized independently.

There was a drop in residency in this second stage as can be seen from Table 3. This time a band width of 2.6 mm was required on average, an increase of 0.60 mm, for 95% residency.

**Table 1:** Residency analysis for Patient 7, Patient 8, Patient 10 and all 52 beams studied (Stage 1).

| **Band Width** | **1 mm** | **2 mm** | **3 mm** | **4 mm** | **5 mm** |
|---|---|---|---|---|---|
| **Patient 7** | 72% | 97% | 99% | 100% | 100% |
| **Patient 8** | 50% | 84% | 96% | 99% | 99% |
| **Patient 10** | 92% | 99% | 100% | 100% | 100% |
| **All 52 beams** | 76% | 95% | 99% | 100% | 100% |

**Table 2***:* Comparison of RMS error results for Patient 7, Patient 8, Patient 10 and all 52 beams studied.

| | Average RMS error | | |
|---|---|---|---|
| | **Stage 1: optimized** | **Stage 2: non-optimized** | **Difference** |
| **Patient 7** | 0.91 mm | 1.16 mm | 0.25 mm |
| **Patient 8** | 1.50 mm | 1.70 mm | 0.20 mm |
| **Patient 10** | 0.58 mm | 0.70 mm | 0.12 mm |
| **All 52 beams** | 0.90 mm | 1.27 mm | 0.37 mm |

**Table 3***:* Residency analysis for Patient 7, Patient 8, Patient 10 and all 52 beams without further optimization (i.e. Stage 2), using Day 1, Beam 1 parameters.

| **Band Width** | **1 mm** | **2 mm** | **3 mm** | **4 mm** | **5 mm** |
|---|---|---|---|---|---|
| **Patient 7** | 62% | 91% | 99% | 100% | 100% |
| **Patient 8** | 42% | 77% | 94% | 99% | 99% |
| **Patient 10** | 85% | 99% | 100% | 100% | 100% |
| **All 52 beams** | 56% | 88% | 98% | 99% | 100% |

### *3.1.1 Detailed analysis of Patients 7, 8, and 10*

The means in Table 2 for patients 7, 8, and 10 were obtained from a subset of data available for those patients. Now we consider all available and reliable data for patients 7, 8, and 10 to determine whether our initial use of a subset of data biased the results shown in Table 2. In Table 4 we show that no such bias exists, because when all available and reliable[3] beam data for Patient 7, Patient 8 and Patient 10 were analyzed, the RMS error and residency results for Patients 7 and 8 calculated from optimized results from all available beam data are close to the values shown in Table 2. The changes in the mean correlation coefficients for Patients 7 and 8 were negligible (about 0.96 in all cases) whereas the mean correlation coefficient for Patient 10 dropped from 0.99 to 0.98 when all available beam data were analyzed, in both Stage 1 and Stage 2.

**Table 4:** Comparison results for all available and reliable data for Patient 7, Patient 8, Patient 10, and all 52 beams studied.

| | **Average RMS error** | | **Band width for 95% residency** | |
|---|---|---|---|---|
| | **Stage 1 optimized** | **Stage 2 non-optimized** | **Stage 1 optimized** | **Stage 2 non-optimized** |
| **Patient 7** | 0.95 mm | 1.36 mm | 2.1 mm | 2.9 mm |
| **Patient 8** | 1.61 mm | 1.87 mm | 3.0 mm | 3.5 mm |
| **Patient 10** | 0.84 mm | 0.95 mm | 1.7 mm | 2.0 mm |
| **All 52 beams** | 0.90 mm | 1.27 mm | 2.0 mm | 2.6 mm |

[3]One beam was omitted for patient 10 because it contained a substantial amount of missing data, and obviously spurious measurements.

*3.1.2 Sensitivity of optimization for different days.*

To test the sensitivity to beam and day parameter optimization, non-optimized results were obtained for Patients 7, 8, and 10 using the parameters obtained from optimizing the Day 3, Beam 2 data, a choice made arbitrarily. The change in the average RMS error and the band width required for 95% residency was small as can be seen from Table 5 below, and there was no change in the mean correlation results. This would indicate that different parameter sets obtained from different beams for the same patient give adequate results.

**Table 5:** Results for all available and reliable data for Patient 7, Patient 8, and Patient 10 optimized using Day 3, Beam 2 parameters.

| | Average RMS error | Mean correlation coefficient | Band width for 95% residency |
|---|---|---|---|
| **Patient 7** | 1.17 mm | 0.96 | 2.5 mm |
| **Patient 8** | 1.91 mm | 0.96 | 3.6 mm |
| **Patient 10** | 0.95 mm | 0.98 | 2.0 mm |

*3.4 Irregular beam data*

The model was capable of dealing with data containing irregularities such as missing data covering several seconds (poor tracking of the abdominal motion), spikes in the amplitude of tumor motion and baseline drift. Figure 5 shows an example of the model applied to data containing some initial baseline drift of the breathing signal, a period of missing data, and an unusual rapid variation in amplitude, or "spike". These data problems are important considerations for modeling lung tumor motion to ensure that errors are not avoidably high.

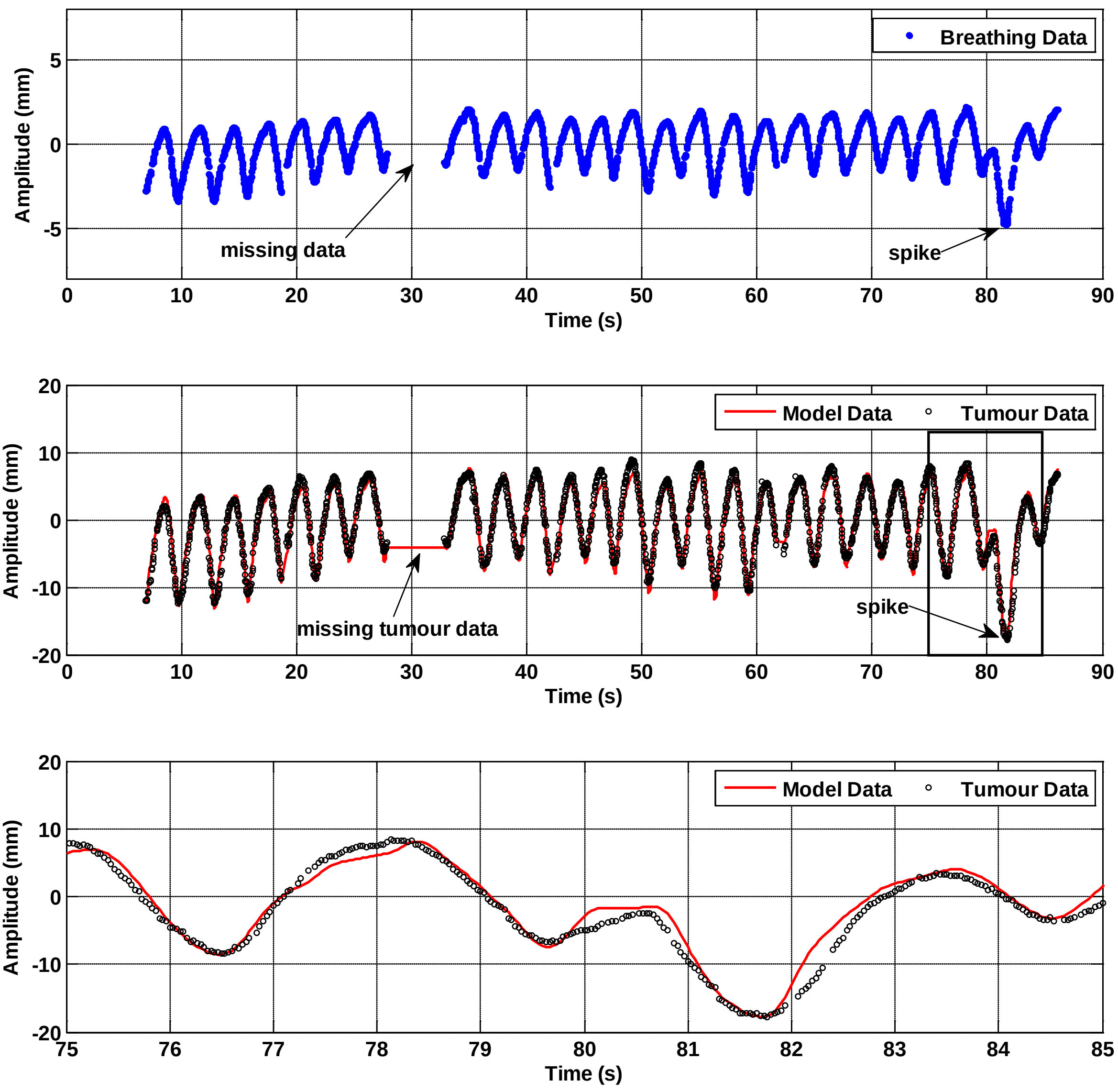

*Figure 5*: Breathing data (top) and tumor data and model output (middle) for Patient 8, Day 2, Beam 1. Labeled are a lengthy period of missing data and a "spike" in the data. The solid line for the model data is an interpolation between points and therefore does not represent actual model output; specifically, the model makes no predictions during periods of missing breathing data. An initial baseline drift is also apparent. The lower figure shows in greater resolution how the model copes with the data spike.

This incomplete data gives a useful clarification of the functioning of the model. The model was not designed to predict tumor motion in the absence of breathing data, only in the absence of tumor data. Thus we see that when the breathing data is missing the model makes no prediction for tumor location. This can be clearly seen in the long period of missing data around 30s, and also in the much shorter periods around 19s and 62s. In all cases, we have plotted an interpolated solid line for the model data for consistency with other plots, although in actual fact the model makes no prediction at those times of missing breathing data.

The RMS error for this beam data is 1.09 mm and the correlation coefficient is 0.98. The band width for 95% residency is 2.1 mm. These results indicate that the model deals well with irregularities in the data. Figure 6 below shows the histogram of residuals and the regression plot for the data presented in Figure 5. The tail of negative residuals in this histogram is long relative to those shown in Figure 4. This is not due to poor systematic performance but rather due to the performance in the short recovery period immediately following the spike in Figure 5, lasting less than one second.

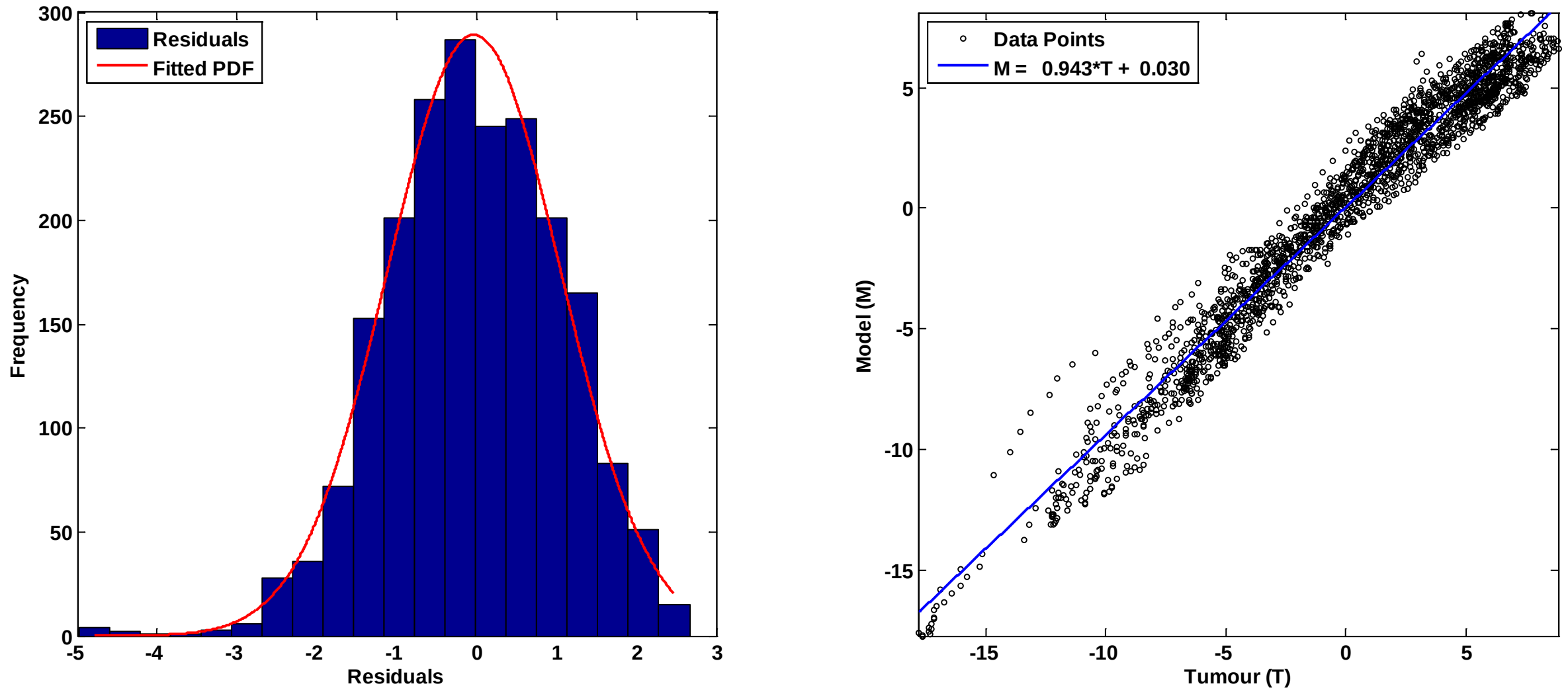

***Figure 6:*** Residual plot (left) and regression plot (right) for Patient 8, Day 2, Beam 1. The mean is -0.030 and standard deviation 1.09, while the correlation coefficient is 0.98.

## 4. Discussion

A modified version of the Wilson & Meyer spring-dashpot model has been developed to correlate abdominal motion with lung tumor motion. In its current implementation the approach is capable of modeling the main component of tumor motion in the SI direction. The model was applied to clinical tumor tracking data from 10 patients treated on the Mitsubishi Real-Time Radiation Therapy system in Sapporo, Japan. In the first stage, the model parameters were optimized for each individual beam in order to determine the goodness of fit for each data set. These values served as a benchmark for further evaluation in Stage 2. In Stage 2, the optimized model parameters for one particular beam were used to estimate the output, i.e. tumor motion, for other beams and treatment days in order to evaluate the trade-offs with regard to optimizing each individual beam separately. The motivation for this was to find out whether it is possible for model parameters to be determined on the first day of treatment or even prior to treatment. The average error determined for Stage 1 was 0.90 ± 0.40 mm, which increased by 0.37 mm in Stage 2. The results were very similar if parameters were used that were initially estimated for other beams for the same patient. This indicates that despite the temporal changes in the abdominal/tumor motion relationship over a treatment course the physical characteristics remain fairly constant, which is reflected in only a small increase in the residual error. This is the advantage of a semi-physical model, such as the one presented here, which aims to model the tissue elasticity as opposed other models which might provide an excellent fit to the input/output data but have no intrinsic relationship to the actual mechanical tissue properties.

To put the results into a more clinically relevant context, hypothetical bounds were calculated so that the modeled point location of the tumor would be within the actual point location of the tumor for 95% of the treatment time. This hypothetical bound would account for the inaccuracies of the modeling and assumes otherwise perfect compensation of the tumor motion, by tumor tracking or respiratory gating, for example. The bound calculated for stage 1 averaged over 52 beams was 2.0 mm and for Stage 2 it was 2.6 mm. To obtain a better understanding of the variability within the patients, three representative patients were further analyzed, patients 7, 8 and 10, corresponding to a residual error of 0.95 mm ('mean'), 1.61 mm (poor') and 0.84 mm ('excellent') in Stage 1. The residual errors showed a largely Gaussian distribution with minimal offset from the mean indicating that the model did not result in systematic model output errors. The calculated error bounds based on the 95% inclusion criteria calculated for the selected patients were 2.1, 3.0, 1.7 mm, respectively, which gives an indication of the correlation between the residual modeling errors and the resulting position uncertainty, which ultimately feeds into the calculation of appropriate margins.

One of the useful features of the spring-dashpot system is that it managed to successfully model baseline drifts and irregular tumor motion. It was also capable of quickly restoring accurate model output when data was missing as shown in Figure 5. The determination of optimal modeling parameters did not include a systematic search of the global parameter space; a downhill search algorithm was used for efficiency. It remains to be investigated if a stochastic search technique, e.g. simulated annealing or approximate Bayesian computation, is required to remove dependence on the initial parameter estimates, but this will be at the expense of higher computational costs. For consistency the initial starting condition for the search was kept constant but it was found that the nominal values of the optimized parameters diverged if different starting conditions were used (data not shown). However, it was found that the resulting residuals were almost identical, which indicates that the solution space is relatively flat. Therefore it was not considered critical but remains an area of further investigation.

For clinical implementation, several extensions to the current modeling process are needed. One is the extension to a full three dimensional system which also considers the minor axes of tumor motion (AP and lateral) for the sake of completeness for the proportion of patients where this motion is non-negligible. Also, ideally the system should be predictive, as there is a finite time required for the treatment machinery (couch[8] or multi-leaf collimator[12]) to adjust to the determined change in position. Ultimately, any algorithm has limitations and is dependent on the quality of the input data and therefore the output of the model should be tested and verified with independent means in a clinical setting.

## 5. Conclusion

A semi-physical spring-dashpot model to correlate breathing to tumor motion in the superior-inferior direction has been presented and applied to clinical tracking data. Optimized model parameters were found to be robust and transferrable to different beams on the same day and consecutive days. Day-to-day variations in the agreement between the model output and the measured data were small, indicating that the model parameters may be determined prior to or on the first day of treatment and then used throughout the course of treatment. The semi-physical nature of the model enabled it to deal with irregularities in the data such as baseline drifts, phase shifts and amplitude and frequency variations. Further work will address the expansion of the model to include all three dimensions and experimental testing and verification of the model output in a clinical setting.

## Acknowledgements

We would like to thank Dr. Seiko Nishioka and Dr. Hiroki Shirato, for the clinical data provided from the Nippon Telegraph and Telephone Corporation Hospital in Sapporo, Japan. Thanks also to the Canterbury Medical Research Foundation for financial support for Alicia Cavan.